%% file: arxiv_version.tex
\documentclass[aps, prl, twocolumn, superscriptaddress, showpacs]{revtex4-1}
\usepackage{amsmath}
\usepackage{amsfonts}
\usepackage{amssymb}
\usepackage{graphicx}
\usepackage{epstopdf}
\usepackage{color}

\newcommand{\si}[0]{$^{29}$Si}

\newcommand{\p}[0]{$^{31}$P}

\begin{document}
\title{Microscopic control of $^{29}$Si nuclear spins near phosphorus donors in silicon}
\date{\today}
\author{J. J\"{a}rvinen}
\email{jaanja@utu.fi}
\affiliation{Wihuri Physical Laboratory, Department of Physics and Astronomy, University of Turku, 20014 Turku, Finland}
\author{D. Zvezdov}
\affiliation{Wihuri Physical Laboratory, Department of Physics and Astronomy, University of Turku, 20014 Turku, Finland}
\affiliation{Institute of Physics, Kazan Federal University, Russia}
\author{J. Ahokas}
\author{S. Sheludyakov}
\author{O. Vainio}
\author{L. Lehtonen}
\author{S. Vasiliev}
\affiliation{Wihuri Physical Laboratory, Department of Physics and Astronomy, University of Turku, 20014 Turku, Finland}
\author{Y. Fujii}
\author{S. Mitsudo}
\author{T. Mizusaki}
\affiliation{Research Center for Development of Far-Infrared Region, University of Fukui, 3-9-1 Bunkyo, Fukui 910-8507, Japan}
\author{M. Gwak}
\author{SangGap Lee}
\affiliation{Division of Materials Science, Korea Basic Science Institute, 169-148 Gwahak-ro, Yuseong-gu, Daejeon 305-806, Korea}
\author{Soonchil Lee}
\affiliation{Department of Physics, Korea Advanced Institute of Science and Technology, 291 Daehak-ro, Yuseong-gu, Daejeon 305-701, Korea}
\author{L. Vlasenko}
\affiliation{A. F. Ioffe Physico-Technical Institute, Russian Academy of Sciences, 194021 St. Petersburg, Russia}

\begin{abstract}
We demonstrate an efficient control of \si{} nuclear spin orientation for specific lattice sites near \p{} donors in silicon crystals at temperatures below 1 K and in high magnetic field of 4.6 T. Excitation of the forbidden electron-nuclear transitions leads to a pattern of narrow holes and peaks in the ESR lines of $^{31}$P. The pattern originates from dynamic polarization the \si{} nuclear spins near the donors via the solid effect. This method can be used for initialization of qubits based on \si{} nuclear spins in the all-silicon quantum computer. In comparison, polarization of \si{} performed by pumping the allowed ESR transitions, did not create any patterns. Instead, a single narrow spectral hole was burnt in the ESR line. The difference is explained by a rapid spin diffusion during the microwave pumping of the allowed transitions.
\end{abstract}


\maketitle

Nuclear spins are among the best candidates for qubits of a quantum computer (QC). Long coherence times and well known magnetic resonance techniques to control and read-out the spin states are the main arguments behind the Kane’s suggestion \cite{Kane1998} of utilizing nuclear spins of phosphorus donors in silicon (Si:P) for this purpose. However, a practical realization of this idea meets with the difficulties of manufacturing complicated nano-structures for manipulating and detecting single spins. Another approach relies on the idea of utilizing large ensembles of identical spins operating coherently, which greatly enhances the net response of the system. Thus a successful 12 qubit operations were realized using nuclear magnetic resonance (NMR) of molecules in liquids \cite{Negrevergne2006}. Different nuclei of the molecules can be addressed because of tiny differences in their resonance frequencies caused by the chemical shifts. A similar approach, also based on large spin ensembles, utilizes spectral holes in inhomogeneously broadened spectral lines \cite{Shahriar2002}. In this case spectrally resolved spin packets with different resonance frequencies are selectively addressed. 

Spin dynamics of P donors in silicon is strongly influenced by the spin-$1/2$ $^{29}$Si nuclei located inside the relatively disperse electron cloud of the donor. For normal isotopic composition there are about 70 such nuclei. Interactions of P electron with these nuclei leads to a loss of coherence of the electron spin. Therefore, the \si{} nuclei were considered as a nuisance and substantial research efforts were directed to the studies of isotopically purified silicon \cite{Simmons2011, Tyryshkin2012, Buch2013}. However, the spins of \si{} nuclei can be also used as qubits \cite{Ladd2002, Pla2014}. Several \si{} nuclei having different  interactions with the donor electron may be addressed selectively by RF excitation. The number of qubits is then defined by the amount of spectrally resolved \si{} nuclei inside the donor electron cloud and may exceed several tens. One may utilize ensembles of these qubits for building an all-silicon QC, in a manner similar to that of the liquid state NMR QC \cite{Negrevergne2006}.

In this work we demonstrate a simple and efficient way to control and read-out \si{} nuclear spins in specific lattice sites near P donors. We found that pumping the forbidden electron-nuclear spin transitions leads to unexpected changes in the shape of phosphorus ESR lines. We observed patterns of narrow spectral holes and peaks corresponding to polarization of \si{} in definite places within the donor electron cloud. The \si{} dynamic nuclear polarization (DNP) is created via the resolved solid effect (SE), and the locations and amplitudes of the holes and peaks coincide well with the calculations.  Polarization of \si{} via the Overhauser effect (OE) by pumping the allowed ESR transition, leads to a burning of a single narrow spectral hole. The absence of patterns in this case is explained by a fast nuclear spin diffusion during pumping. The polarized states of \si{} survived for many hours. We explain the long life time by locking the \si{} nuclear spins inside a spin diffusion barrier around the donors.

We used a sample of phosphorus doped ($n(\text{P})\approx6.5\times10^{16}$ cm$^{-3}$) natural silicon (4.7 \% $^{29}$Si), cut to a $2\times 2$ mm wide and 70 $\mu$m thick square with the $\langle 111 \rangle$ crystal axis perpendicular to the sample surface and parallel to the static magnetic field. The sample was glued on the flat mirror of an open Fabry-Perot resonator \cite{jarvinen2014} connected to a cryogenic heterodyne EPR spectrometer operating at a frequency of 128 GHz \cite{SpectrometerRSI} in a magnetic field of 4.6 T. The spectrometer provides high sensitivity at ultra-low excitation powers of several pW without field or frequency modulation, using a synthesized, nearly monochromatic, mm-wave source \cite{VDI}. This provides sensitive detection of undistorted absorption and dispersion ESR lineshapes with high resolution. For pumping the ESR transitions the highest available excitation power of 400 nW was used. Sample cell was cooled to temperatures below 1 K by a dilution refrigerator.

\begin{figure}
\includegraphics[width=\columnwidth]{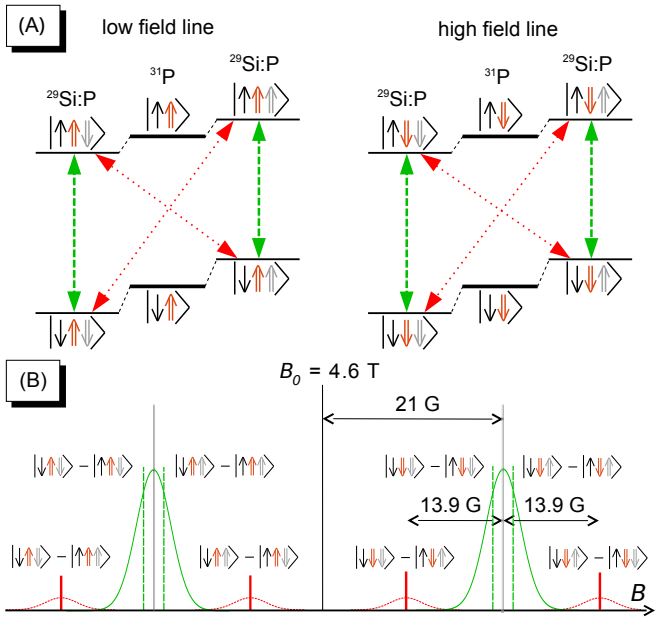}
\caption{\label{transitions} (Color online) (A) Level diagram and ESR transitions of electron spin interacting with a \p{} and \si{} nuclear spins in strong magnetic field. Arrows in the brackets denote the donor electron, \p{}, and \si{} nuclear spin states. Dashed arrows mark the allowed ESR transitions, dotted arrows - forbidden transitions with simultaneous spin flips of electron and \si{} nuclear spins.(B) Schematic of relevant transitions in Si:P ESR spectrum. The main phosphorus ESR transitions are illustrated with long green lines and the forbidden electron-\si{} spin transitions with the short red lines. }
\end{figure}

The well known ESR spectrum of $^{31}$P in silicon consists of two lines, separated by 42 G due to the hyperfine interaction of the donor electron with its own nucleus. The superhyperfine interactions with \si{} nuclei lead to further level splittings and to $\approx4$ G broadening of the \p{} ESR line (Fig. \ref{transitions}A). In strong magnetic fields the forbidden transitions are well separated from the allowed ones (Fig. \ref{transitions}B), and one may realize resolved SE or OE. In this work we shall consider transitions for the \si{}:P system. The nuclear spin of the donor remains unchanged.

\begin{figure}
\includegraphics[width=\columnwidth]{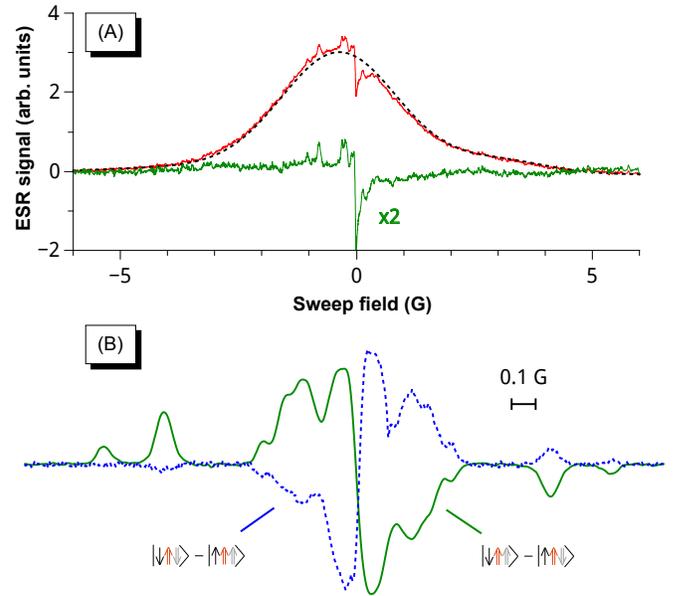}
\caption{\label{solid} (Color online) (A) Spectrum of the solid effect after exciting $\left|\downarrow \Uparrow \Uparrow\right\rangle \rightarrow \left|\uparrow \Uparrow \Downarrow\right\rangle$ transition at 200 mK for 36 min. The dashed black line is a Gaussian fit to the line. The red solid line with the burned hole is the spectrum just after the DNP with solid effect. The green curve with flat background is subtracted spectrum multiplied by two. (B) Solid effect ESR spectrum of transitions ($\left|\downarrow \Uparrow \Uparrow\right\rangle \rightarrow \left|\uparrow \Uparrow \Downarrow\right\rangle$) (exited for 115 min) and ($\left|\downarrow \Uparrow \Downarrow\right\rangle \rightarrow \left|\uparrow \Uparrow \Uparrow\right\rangle$) (exited for 140 min) plotted correspondingly with green solid and blue dashed line.}
\end{figure}

We performed SE DNP of the \si{} by pumping one of the four forbidden transitions located at $\pm13.9(1)$ G from the ESR lines of \p{} (Fig. \ref{transitions}B). The forbidden transitions are too weak to be detected directly, but the result of the pumping can be seen well in the ESR line. In Fig. \ref{solid}A the low field line is shown after 36 min of pumping the $\left|\downarrow \Uparrow \Uparrow\right\rangle \rightarrow \left|\uparrow \Uparrow \Downarrow\right\rangle$ transition at temperature of 200 mK. Subtraction of the undisturbed ESR line reveals an anti-symmetric pattern of peaks and holes. A similar experiment performed on the $\left|\downarrow \Uparrow \Downarrow\right\rangle \rightarrow \left|\uparrow \Uparrow \Uparrow\right\rangle $ transition located at -13.9 G from the low field line, produced an inverted pattern. In Fig. \ref{solid}B we present both patterns obtained after 200 signal averages for improving the signal-to-noise. At temperatures below 1 K the patterns can be studied for many hours without any significant change in their shape. Similar results were obtained after pumping the forbidden transitions near the high field line of $^{31}$P. The resulting pattern, as we will show below, is caused by the DNP of \si{} in certain lattice sites. This is a new and impressive feature of the resolved SE, especially taking into account the relatively weak saturation of the forbidden transitions with small (400 nW) ESR power.

For the explanation of the observed patterns we consider a donor electron spin interacting with N \si{} nuclei. Energy levels can be found using the following Hamiltonian
\begin{equation}\label{hamiltonian}
H=-g_e \mu_b B_0 S_z + \sum_{k=1}^N  (-g_n \mu_n B_0 I_{z,k} - a_k S_z I_{z,k} - S\cdot T_k \cdot I_k).
\end{equation}
Here $g_e=-1.99875$ and $g_n=-1.11058$ are the P electron and $^{29}$Si nuclear g factors, $\mu_b$ and $\mu_n$ are the Bohr and nuclear magnetons,
$S$ and $I$ are the electron and nuclear spin operators, $a_k$ is the Fermi contact interaction and $T_k$ is a symmetric tensor of the anisotropic hyperfine interaction. The index $k$ labels lattice sites randomly occupied by $^{29}$Si nuclei. Using the following approximations $|g_e \mu_b B| \gg |g_n \mu_n B| \gg |a_k| \gg
|T_k|$, and neglecting the anisotropic hyperfine interaction the frequencies of the allowed transitions are (ref. \cite{Feher1959, Poole1972})
\begin{equation}\label{allowed}
h f_{x} = g_e \mu_b B_0 - \sum_{k=1}^N m_k \frac{a_k}{2},
\end{equation}
with $m_k= 4 m_s m_{I,k}=\pm 1$, where $m_s$ and $m_{I,k}$ are the magnetic quantum numbers of the electron and nuclear spin in the site $k$. Each donor in the sample has a specific number and configuration of surrounding \si{}, which we label with $N$ and $x$. This leads to a spread of the energy levels and inhomogeneous broadening of the allowed and forbidden transitions. Let's consider a flip-flop transition with the flop of a single \si{} spin in a lattice site $l$. The frequency is
\begin{equation}\label{forbidden}
h f^{'}_{x,l} = g_e \mu_b B_0 + g_n \mu_n B_0  - \sum_{k=1}^N m_k \frac{a_k}{2}-\frac{a_l}{2}.
\end{equation}
However, the difference between the forbidden and allowed transition frequencies does not depend on $x$, or on the states of the nuclei $k\neq l$ which are not involved in the transition. Indeed, we find from Eqs. (\ref{allowed}) and (\ref{forbidden}):
\begin{equation}\label{separation}
h (f^{'}_{x,l}-f_{x}) =g_n \mu_n B_0 -\frac{a_l}{2}.
\end{equation}
The distance between the allowed and forbidden transition is solely a function of the hyperfine constant for the lattice site $l$ where the nuclear \si{} spin is flipped. Following the spin transfer path for SE via the flip-flop transition  and subsequent electron relaxation (see Fig. \ref{transitions}A) we expect to find a peak on the left and a hole on the right both separated from the center of the pattern by $\pm a_l /2$. For the flip-flip transition the pattern is inverted, which agrees well with our observations (Fig. \ref{solid}).

\begin{figure}
\includegraphics[width=\columnwidth]{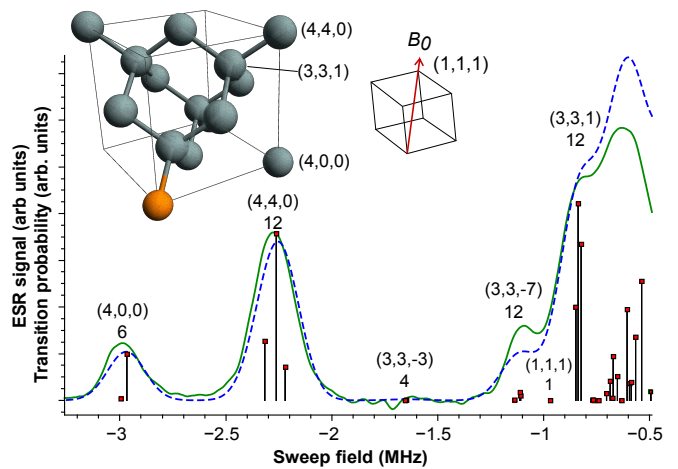}
\caption{\label{fig3b}(Color online) Spectrum of the solid effect (green) and calculated spectrum (dashed blue). The calculated transition probabilities are marked with the black vertical lines. The small splittings of the transitions are due to anisotropic hyperfine interaction. Above the peaks in the spectrum the coordinates of the lattice sites together with the number of lattice sites belonging to each line are marked.}
\end{figure}

For calculating the intensities of the flip-flip and flip-flop transitions we use the Hamiltonian (\ref{hamiltonian}). Now we take into account the anisotropic hyperfine interaction \cite{Feher1959, Hale1969, Ivey1975} because it slightly mixes the spin states. However, we can safely neglect the off-diagonal terms of the Hamiltonian connecting different $S_z$ states \cite{Poole1972} and direct interactions between \si{} spins. Evaluating the transition probabilities for each lattice site $k$ we find predictions for the SE patterns \cite{supplement} which are in good agreement with our observations (Fig. \ref{fig3b}). 

Next, we studied DNP of \si{} by pumping the allowed ESR transitions. Saturating at some point of an inhomogeneously broadened ESR line typically results in a spectral hole caused by the depolarization of electron spins. After removing the excitation the hole vanishes due to electron spin relaxation or diffusion. However, the electron spin polarization may be transferred to the nuclei of surrounding \si{} via the DNP processes. This will also result in a spectral hole which will fade out very slowly, with the rate determined by the nuclear relaxation. Hole burning experiments on Si:P performed earlier in low magnetic fields and at high temperatures\cite{Feher1959, Feher1959a, Marko1970} have not clarified the actual mechanisms of the hole burning and relaxation.

\begin{figure}
\includegraphics[width=\columnwidth]{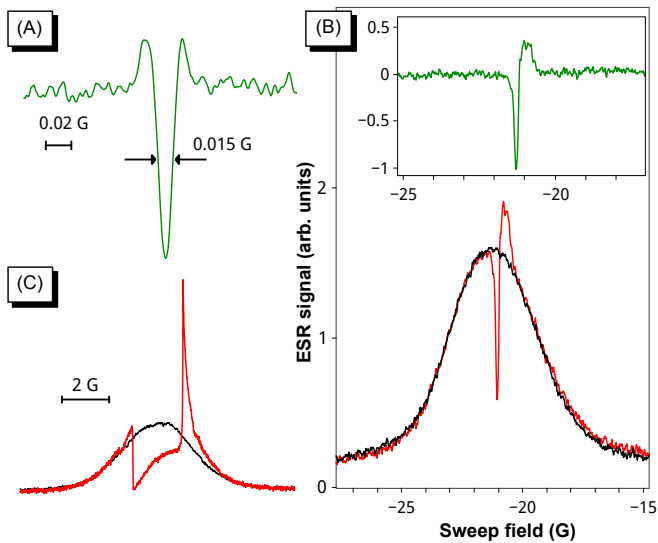}
\caption{\label{spectra} (Color online) (A) The narrowest burned hole in the low field line. (B) Low field ESR line of Si:P recorded before and after burning a hole. The inset shows the hole without the background signal. (C) Lineshape after applying ESR pumping with 3 MHz FM modulation at 40 nW power to the center of the low field line for 10 min.}
\end{figure}

Results of pumping for several minutes at the center of the low-field ESR line are presented in Fig. \ref{spectra}A. A single hole appeared in the spectrum accompanied with two narrow peaks on both sides. The observed $\approx15$ mG width of the hole in reasonable agreement with the transversal relaxation rate of the electron spin $T_{2}^{-1}\approx 40$ kHz \cite{Abe2010}. Next, we pumped using FM modulated source with 100 Hz rate and 100 kHz deviation. This also created a hole and a broader peak at the high field side of the hole (Fig. \ref{spectra}B). In either of the cases above we have not observed any pattern of holes and peaks, contrary to the SE DNP experiments. Finally, we performed pumping with the FM deviation of $\pm 3$ MHz. This resulted in a "burnt" window (Fig. \ref{spectra}C) with a strong peak near the right edge and much weaker peak near the left edge. 

At temperatures below 0.5 K the results of the ESR line pumping: holes, peaks and windows faded out very slowly, with the characteristic time of several hours, much longer than the electron spin-lattice relaxation time $T_1\approx 0.2$ s. We were able to destroy the patterns and holes created by SE and OE by applying excitation at the NMR frequency of \si{} to the RF coil located near the Fabry-Perot resonator \cite{jarvinen2014}. All the observations above show unambiguously that the pumping of the spin packets leads, not only to their saturation, but also to redistribution of the $^{29}$Si nuclear spin orientations near the phosphorus donors. 

For the explanation of the hole burning experiments we take into account the \si{} nuclear spin diffusion. This process has a high rate $\gtrsim10^{-14}$ cm$^2$/s in bulk silicon crystal \cite{Hayashi2008} but is strongly suppressed inside a so-called diffusion barrier created by the dipole fields of the electron spins near donors \cite{Khutsishvili1969}. The 10 nm size of the diffusion barrier in natural silicon is much larger than the span of the donor electron cloud. A weak saturation of the forbidden transitions in the SE experiments does not destroy the diffusion barrier. Therefore, the \si{} polarization created with the SE remains at the original lattice sites for a long time. 

In the OE experiment, on the contrary, the allowed transitions are fully saturated and thus the electron spins are depolarized. This quenches the diffusion barrier \cite{Khutsishvili1969} and allows rapid spreading of the \si{} polarization. The OE polarization, arising near the donor where the \si{} have the strongest interaction with the donor electron, is rapidly conducted away from the donor to the nuclei with weaker interactions. This explains why we do not observe patterns of peaks and holes in the OE DNP experiments. Once the \si{} nuclear spin is flipped, the donor nearby gets out of resonance. Then the spin diffusion barrier develops around the donor with a characteristic time $T_1 \approx 0.2$ s. The \si{} polarization may propagate over a distance of few lattice constants during $T_1$ before getting frozen. Such redistribution of the \si{} nuclear polarization under pumping may lead to appearance of the sharp peaks near the burnt hole, as is observed in our experiments (Fig. \ref{spectra}). During a window pumping the donors are not getting out of resonance, but rather remain inside the window and are re-pumped again. That is why the peaks at the sides of the window are much stronger.

Next we explain why the stronger peak appears at the high field (right) side of the burnt window. During the OE pumping we excite equal amounts of the allowed $\left|\downarrow \Uparrow \Downarrow\right\rangle \rightarrow \left|\uparrow \Uparrow \Downarrow\right\rangle$ and $\left|\downarrow \Uparrow \Uparrow\right\rangle \rightarrow \left|\uparrow \Uparrow \Uparrow\right\rangle$ transitions if the \si{} spins are unpolarized. Because the flip-flop relaxation is usually faster than the flip-flip \cite{Jeffries1960}, the  OE  will enhance population of the $\left|\downarrow \Uparrow \Uparrow\right\rangle$ state. Since the $\left|\downarrow \Uparrow \Uparrow\right\rangle \rightarrow \left|\uparrow \Uparrow \Uparrow\right\rangle$ occur in higher sweep field, such re-distribution of the spin states leads to the increase of the signal on the right side from the pumping position. Calculating the areas of the peaks on the right and left of the pumped window in Fig. \ref{spectra}C we evaluate the ratio of the flip-flop and the flip-flip relaxation rates $\approx8$. This is substantially smaller than for the similar flips of \p{} and its own nucleus \cite{jarvinen2014}, and may be caused by a stronger anisotropy of superhyperfine interactions.

Once the \si{} nuclear spins are polarized in the distinct shells, they can be further manipulated by applying resonant microwave pulses at the frequencies of the corresponding NMR transitions. Coherent superpositions of spin-up and down (qubit) states for the ensembles of \si{} nuclei can be created. The NMR manipulations will lead to the shifts of the positions of the peaks in the ESR line, and thus the projections of the spin up and down states can be measured. In the patterns presented in Fig. 3 two locations are well resolved: the (4,0,0) and (4,4,0) sites, having NMR frequencies f(4,0,0)$\approx$41.88 MHz, f(4,4,0)$\approx$41.15 MHz. The separation between the peaks corresponding to the different lattice sites can be changed by varying the orientation of the Si crystal in magnetic field. It might be beneficial using isotopically enriched \si{} crystals for getting larger ensembles of nuclei and improving signal-to-noise.

In a conclusion we have demonstrated that the solid effect DNP can be used for polarizing $^{29}$Si nuclei located in specific lattice sites near phosphorus donors. Spin diffusion does not destroy the polarization inside the diffusion barriers. In the OE DNP the diffusion barrier is quenched due to saturation of allowed electronic transitions which leads to a rapid spread of polarization in the bulk of the silicone crystal. The SE DNP  demonstrated in our work, opens up possibilities of using ensembles of \si{} nuclear spins for quantum information storage and processing with the read-out based solely on ESR technique.

\begin{acknowledgments}
We would like to thank K. Itoh and S. Lyon for useful discussions. We acknowledge the funding from the Wihuri Foundation and the Academy of Finland grants No. 260531 and 268745. S. S. thanks UTUGS for support. L.V. acknowledges partial support of the Government of Russia, proj. 14.Z50.31.0021, M.G. and S.-G.L. acknowledge KBSI Grant T34412 for funding support.
\end{acknowledgments}

\bibliography{silicon}
\bibliographystyle{apsrev4-1}

\pagebreak

\input{arxiv_supplementary}

\end{document}

%% file: arxiv_supplementary.tex

\pagebreak
\onecolumngrid
\section*{Supplementary material}

The Hamiltonian describing superhyperfine level splitting of an electron
spin $S$ interacting with a $^{29}$Si nuclear spin $I_{k}$ is

\begin{equation}
H=H_{0}+H'=-g_{e}\mu_{b}B_{0}S_{z}-g_{n}\mu_{n}B_{0}I_{z,k}-a_{k}S \cdot I_{k}-S\cdot T_{k}\cdot I_{k},\label{eq:h0}
\end{equation}
where $H_{0}=-g_{e}\mu_{b}B_{0}S_{z}-g_{n}\mu_{n}B_{0}I_{z,k}-a_{k}S \cdot I_{k}$ is the isotropic part, and
$H_{k}=S\cdot T_{k}\cdot I_{k}$ is the anisotropic part of the Hamiltonian with the gyromagnetic factors $g_{e}$ and $g_{n}$ being negative. The isotropic superhyperfine constant
$a_{k}$ is given by 
\[
a_{k}=\frac{2\mu_{0}}{3}g_{e}\mu_{b}{}_{s}g_{n}\mu_{n}\hbar^{2}|\psi(r_{k})|^{2},
\]
where $\psi$ is the electron wavefunction. The minus sign before
$a_{k}$ in the Eq. ($\ref{eq:h0}$) takes into account the negative
$g_{n}$ of the $^{29}$Si making the values of the superhyperfine
constant positive. The anisotropic superhyperfine interaction tensor
$T_{k}$ for the electron with the nuclei in position $r_{k}=(r_{k,x},r_{k,y},r_{k,z})$
is 
\[
T_{k}(i,j)=\frac{\mu_{0}}{4\pi}g_{e}\mu_{b}g_{n}\mu_{n}\hbar^{2}\left\langle \psi\left|\frac{r_{k}{}^{2}\delta_{ij}-3r_{k,i}r_{k,j}}{r_{k}^{5}}\right|\psi\right\rangle .
\]
 In the matrix form the spin projections are defined as

\[
\psi=\left(\begin{array}{c}
\uparrow\Uparrow\\
\uparrow\Downarrow\\
\downarrow\Uparrow\\
\downarrow\Downarrow
\end{array}\right),
\]
where $\downarrow$ is the electron spin and $\Downarrow$ is the
$^{29}$Si nuclear spin. The isotropic part of the Hamiltonian (index $k$ is omitted)
is

\begin{equation}
H_{0}=\left(\begin{array}{cccc}
-\frac{a}{4}-\frac{g_{e}\mu_{B}B_{0}}{2}-\frac{g_{n}\mu_{n}B_{0}}{2} & 0 & 0 & 0\\
0 & \frac{a}{4}-\frac{g_{e}\mu_{B}B_{0}}{2}+\frac{g_{n}\mu_{n}B_{0}}{2} & -\frac{a}{2} & 0\\
0 & -\frac{a}{2} & \frac{a}{4}+\frac{g_{e}\mu_{B}B_{0}}{2}-\frac{g_{n}\mu_{n}B_{0}}{2} & 0\\
0 & 0 & 0 & -\frac{a}{4}+\frac{g_{e}\mu_{B}B_{0}}{2}+\frac{g_{n}\mu_{n}B_{0}}{2}
\end{array}\right),
\label{isotropicH}
\end{equation}
which gives the energy eigenvalues

\[
E_{0}=\left(\begin{array}{c}
-\frac{a}{4}-\frac{g_{e}\mu_{B}B_{0}}{2}-\frac{g_{n}\mu_{n}B_{0}}{2}\\
\frac{1}{4}\left(a-2\sqrt{a^{2}+(g \mu_{B}-g_{n}\mu_{n})^{2}B_{0}^{2}}\right)\\
\frac{1}{4}\left(a+2\sqrt{a^{2}+(g \mu_{B}-g_{n}\mu_{n})^2 B_{0}^{2}}\right)\\
-\frac{a}{4}+\frac{g_{e}\mu_{B}B_{0}}{2}+\frac{g_{n}\mu_{n}B_{0}}{2}
\end{array}\right)
\]

\[
\approx\left(\begin{array}{c}
-\frac{a}{4}-\frac{g_{e}\mu_{B}B_{0}}{2}-\frac{g_{n}\mu_{n}B_{0}}{2}\\
\frac{a}{4}-\frac{g_{e}\mu_{B}B_{0}}{2}+\frac{g_{n}\mu_{n}B_{0}}{2}\\
\frac{a}{4}+\frac{g_{e}\mu_{B}B_{0}}{2}-\frac{g_{n}\mu_{n}B_{0}}{2}\\
-\frac{a}{4}+\frac{g_{e}\mu_{B}B_{0}}{2}+\frac{g_{n}\mu_{n}B_{0}}{2}
\end{array}\right).
\]
The separations of the peak patterns in the ESR spectrum are given
by the differences between the flip-flip $\downarrow\Downarrow\rightarrow\uparrow\Uparrow$
or the flip-flop $\downarrow\Uparrow\rightarrow\uparrow\Downarrow$
and the allowed transition $(\downarrow\Downarrow\rightarrow\uparrow\Downarrow,\downarrow\Uparrow\rightarrow\uparrow\Uparrow)$
frequencies. E.g. difference between the allowed and the flip-flop
transition frequency,
\[
\Delta f_{k}=((E_{0}(1)-E_{0}(3))-(E_{0}(2)-E_{0}(3)))=-g_{n}\mu_{n}B_{0}-\frac{a_{k}}{2},
\]
is positive, indicating that the flip-flop transition
is at higher sweep field on the right-hand side of the allowed transitions.

\begin{figure}
\begin{center}
\includegraphics[]{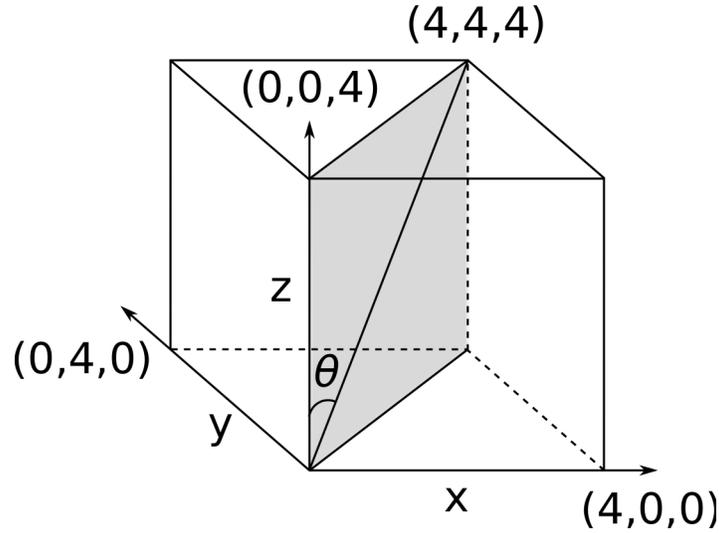}
\end{center}
\caption{Coordinate axis of $T_{k}$ tensor in Si lattice.\label{fig:Coordinate-axis}}
\end{figure}

\begin{table}
\caption{Values of isotropic and anisotropic superhyperfine constants in kHz
units for Si:P for various shells. Class refers to symmetry class of the shell and Nr. refers to the number of lattice sites in the shell. The values are calculated from the refs. \cite{Feher1959,Hale1969,Ivey1975}.\label{tab:Superhyperfine constants}}

\begin{center}

\begin{tabular}{|c|c|c|c|c|c|c|c|c|c|c|}
\hline 
Shell & Site & Class & Nr & $\frac{a}{2}$ & $T_{xx}$ & $T_{yy}$ & $T_{zz}$ & $T_{xy}$ & $T_{xz}$ & $T_{yz}$\tabularnewline
\hline 
\hline 
A & $004$ & 1 & 6 & 2981 & -20.7 & -20.7 & 41.4 & 41.4 & 0 & 0\tabularnewline
\hline 
B & $440$ & 2 & 12 & 2254 & 17 & 17 & -34 & 106.2 & -39.8 & -39.8\tabularnewline
\hline 
C & $33\overline{3}$ & 3 & 4 & 1649 & 0 & 0 & 0 & 5 & -5 & -5\tabularnewline
\hline 
D & $33\overline{7}$ & 2 & 12 & 1117 & -1.8 & -1.8 & 3.6 & 16.6 & 22.2 & 22.2\tabularnewline
\hline 
E & $111$ & 3 & 4 & 270 & 0 & 0 & 0 & 700 & 700 & 700\tabularnewline
\hline 
F & $331$ & 2 & 12 & 840 & -58.2 & -58.2 & 116.4 & -28.2 & -11.8 & -11.8\tabularnewline
\hline 
G & $77\overline{3}$ & 2 & 12 & 764 & -2.5 & -2.5 & 5 & -1.2 & 5 & 5\tabularnewline
\hline 
H & $44\overline{4}$ & 3 & 4 & 689 & 0 & 0 & 0 & 50.6 & -50.6 & -50.6\tabularnewline
\hline 
I & $22\overline{8}$ & 2 & 12 & 685 & 8.7 & 8.7 & -17.4 & 27.8 & 13.8 & 13.8\tabularnewline
\hline 
M & $22\overline{4}$ & 2 & 12 & 612 & 20.3 & 20.3 & -40.6 & 49.6 & 35.6 & 35.6\tabularnewline
\hline 
O & $444$ & 3 & 4 & 598 & 0 & 0 & 0 & 32.6 & 32.6 & 32.6\tabularnewline
\hline 
Q & $115$ & 2 & 12 & 524 & 20.1 & 20.1 & -40.2 & 84 & 22.8 & 22.8\tabularnewline
\hline 
R & $771$ & 2 & 12 & 379 & -10.5 & -10.5 & 21 & 0.2 & 19.4 & 19.4\tabularnewline
\hline 
X & $551$ & 2 & 12 & 317 & -14.7 & -14.7 & 29.4 & 23.6 & 32.2 & 32.2\tabularnewline
\hline 
\end{tabular}

\end{center}
\end{table}

Due to the symmetry of the silicon crystal the lattice sites can be divided into lattice shells $(A,B,C,...)$ for which the $T_{k}$ tensors are not independent. Every lattice site in the crystal belongs to one
of the shells. In addition, each shell can be assigned to one of 4
classes ($S_{1}$, $S_{2}$ , $S_{3}$ and $S_{4}$) depending on
the symmetry of the shell. There are 6 independent components in $T_{k}$
in most general case which is the class $S_{4}$. For all the other
classes only $\leq\text{4 }$components are needed. All the closest
lattice sites belong to the first 3 classes ($S_{1}-S_{3})$. The
symmetry constraints of $T_{k}$ components for each shell are taken
from ref. \cite{Hale1969} and together with the requirement of $\text{Tr}[T]=0$
all the components of $T_{k}$ can be calculated. We present the components of the anisotropic hyperfine tensor
\[
T_{k}=\left(\begin{array}{ccc}
T_{xx} & T_{xy} & T_{xz}\\
T_{xy} & T_{yy} & T_{yz}\\
T_{xz} & T_{yz} & T_{zz}
\end{array}\right)
\]
in Table \ref{tab:Superhyperfine constants} for some of the closest shells.

Now, if $T_{k}$ is known for one nucleus in a shell, the others can
be evaluated by using rotations of the shell class, exchanging any
of the two nuclei in the shell and not influencing the lattice or
the dopant structure \cite{Hale1969}. This can be achieved with a combination of two rotation operators $R_{x}(\theta)$ and $R_{y}(\theta)$ rotating the lattice around x- and y-axis shown in Table \ref{tab:rotations.}.
Finally the whole lattice is rotated so that the $B_{0}$ field is
aligned with the $111$-axis (angle $\theta=54.74^{\circ}$). The
transformed components $T_{k}^{'}$ are then given by

\[
T_{k}^{'}=\text{R}_{x}^{T}[-\theta]\cdot\text{R}_{z}^{T}[-\pi/4]\cdot\text{R}_{i}[\theta_{i}]\cdot\text{R}_{j}[\theta_{j}]\cdot T_{k}\cdot\text{R}_{j}^{T}[\theta_{j}]\cdot\text{R}_{i}^{T}[\theta_{i}]\cdot\text{R}_{z}[-\pi/4]\cdot\text{R}_{x}[-\theta],
\]
where $R_{i}$ and $R_{j}$ are x and y rotations by the angles $\theta_{i}$
and $\theta_{j}$ which depends on the class
of the shell (Table \ref{tab:rotations.}). Now, in the high field aproximation $|T_{ij}|\ll|g_{e}\mu_{B}B_{0}|$ we neglect the components of $T_{k}$ connecting different electron
spin states \cite{Poole1972}. The anisotropic part of the Hamiltonian is then

\begin{equation}
H'=\left(\begin{array}{cccc}
-\frac{T_{zz}^{'}}{4} & -\frac{T_{xz}^{'}}{4}+\frac{iT_{yz}^{'}}{4} & 0 & 0\\
-\frac{T_{xz}^{'}}{4}-\frac{iT_{yz}^{'}}{4} & \frac{T_{zz}^{'}}{4} & 0 & 0\\
0 & 0 & \frac{T_{zz}^{'}}{4} & \frac{T_{xz}^{'}}{4}-\frac{iT_{yz}^{'}}{4}\\
0 & 0 & \frac{T_{xz}^{'}}{4}+\frac{iT_{yz}^{'}}{4} & -\frac{T_{zz}^{'}}{4}
\end{array}\right).\label{eq:full H}
\end{equation}

\begin{table}
\caption{Rotations of the symmetry in different shell classes.\label{tab:rotations.}}

\begin{center}
\begin{tabular}{|c|c|c|c|}
\hline 
Nr. & $S_1$ & $S_2$ & $S_3$\tabularnewline
\hline 
\hline 
1 & $0$ & $0$ & $0$\tabularnewline
\hline 
2 & $x(\pi)$ & $x(\pi)$ & $x(\pi)$\tabularnewline
\hline 
3 & $y(\frac{\pi}{2})x(\frac{\pi}{2})$ & $y(\pi)$ & $y(\pi)$\tabularnewline
\hline 
4 & $y(\frac{\pi}{2})x(-\frac{\pi}{2})$ & $y(\pi)x(\pi)$ & $y(\pi)x(\pi)$\tabularnewline
\hline 
5 & $x(\frac{\pi}{2})y(\frac{\pi}{2})$ & $y(\frac{\pi}{2})x(\frac{\pi}{2})$ & \tabularnewline
\hline 
6 & $x(\frac{\pi}{2})y(-\frac{\pi}{2})$ & $y(-\frac{\pi}{2})x(-\frac{\pi}{2})$ & \tabularnewline
\hline 
7 &  & $x(\frac{\pi}{2})y(\frac{\pi}{2})$ & \tabularnewline
\hline 
8 &  & $x(-\frac{\pi}{2})y(\frac{\pi}{2})$ & \tabularnewline
\hline 
9 &  & $x(\frac{\pi}{2})y(-\frac{\pi}{2})$ & \tabularnewline
\hline 
10 &  & $x(-\frac{\pi}{2})y(-\frac{\pi}{2})$ & \tabularnewline
\hline 
11 &  & $y(-\frac{\pi}{2})x(\frac{\pi}{2})$ & \tabularnewline
\hline 
12 &  & $y(\frac{\pi}{2})x(-\frac{\pi}{2})$ & \tabularnewline
\hline 
\end{tabular}

\end{center}
\end{table}

\begin{table}
\caption{Superhyperfine energies including anisotropic interaction and flip-flop transition probabilities $\Gamma_{k}$ in Si:P when the magnetic field is parallel to the $[111]$-axis of the
crystal. Only the shells which are used in the calculated spectrum on Fig. 3 in the main article are listed.  Transition probabilities are given relative to the allowed transition probabilities.} \label{tab:transition}

\begin{center}

\begin{tabular}{|c|c|r|r||c|c|r|r|}
\hline 
Shell & Nr. & Energy (kHz) & $\Gamma_{k}(10^{-9})$ & Shell & Nr. & Energy (kHz) & $\Gamma_{k}(10^{-9})$\tabularnewline
\hline 
\hline 
A & 3 & 2994.8 & 16 & H & 1 & 739.6 & 0\tabularnewline
\hline 
A & 3 & 2967.2 & 398 & H & 3 & 672.1 & 376\tabularnewline
\hline 
B & 3 & 2316.0 & 509 & I & 3 & 703.5 & 59\tabularnewline
\hline 
B & 3 & 2262.9 & 1434 & I & 3 & 685.1 & 168\tabularnewline
\hline 
B & 6 & 2218.6 & 288 & I & 6 & 675.7 & 21\tabularnewline
\hline 
C & 1 & 1654.0 & 0 & M & 3 & 652.3 & 206\tabularnewline
\hline 
C & 3 & 1647.4 & 4 & M & 3 & 604.8 & 784\tabularnewline
\hline 
D & 3 & 1137.3 & 4 & M & 6 & 595.5 & 144\tabularnewline
\hline 
D & 6 & 1111.5 & 72 & Q & 3 & 567.2 & 542\tabularnewline
\hline 
D & 3 & 1107.7 & 41 & Q & 3 & 536.8 & 1025\tabularnewline
\hline 
E & 1 & 970 & 0 & Q & 6 & 496.0 & 78\tabularnewline
\hline 
E & 3 & 35.3 & 71970 & O & 1 & 630.6 & 0\tabularnewline
\hline 
F & 6 & 849.4 & 803 & O & 3 & 587.1 & 156\tabularnewline
\hline 
F & 3 & 838.4 & 1692\tabularnewline
\cline{1-4} 
F & 3 & 822.7 & 1340\tabularnewline
\cline{1-4}
\end{tabular}

\end{center}
\end{table}
This Hamiltonian mixes the undistorted states of the isotropic Hamiltonian $H_0$ (Eq. (\ref{isotropicH})), thus making the flip-flip and flip-flop transitions allowed. The probabilities of these transitions excited by the
transversal microwave field are $\propto B_{1}\cdot S$, which for linear
excitation is $B_{1}S_x$. Using the Fermi's golden rule we get
\[
P_{ij}=C_1\left|\left\langle \phi_{i}\right|S_{x}\left|\phi_{j}\right\rangle \right|^{2},
\]
where $C_1$ is a constant dependent on the cavity coupling and the excitation power. In Table \ref{tab:transition} we list the flip-flop transition probabilities normalized to the allowed transition probabilities as:
\[
\Gamma_{k}=\frac{\left|\left\langle \uparrow\Downarrow\right|S_{x}\left|\downarrow\Uparrow\right\rangle \right|_{k}^{2}}{\left|\left\langle \uparrow\Uparrow\right|S_{x}\left|\downarrow\Uparrow\right\rangle \right|_{k}^{2}}.
\]
The calculations give equal values for the flip-flip and flip-flop
transition probabilities. 

The observed spectrum (see Fig. 3 in main text) after exciting the flip-flop or flip-flip transition
depends on the strength and duration of the microwave excitation $B_{1}$.
During the excitation each hole and peak in the pattern grows according
to its transition rate $\Gamma_{k}$. A following fitting function
was used to compare the measured pattern to the calculated transition
probabilities:

\begin{equation}
S(\Delta f)=A\sum_{k=1}^{N}(1-\exp(-\Gamma_{k}t_{p}))G_{k}(\Delta f),\label{eq:fit}
\end{equation}
where $G_{k}(\Delta f)$ is the lineshape function, $t_{p}$ is the
effective ESR pumping time which depends on the excitation field strength
and $C$ is the linewidth. A Gaussian function 
\[
G_{k}(\Delta f)=\frac{1}{C\sqrt{2\pi}}\exp(-\frac{1}{2C^{2}}(\Delta f\pm \frac{a_{k}}{2})^{2})
\]
was used as the intrinsic lineshape. The +(-) in the equation is used for fitting the peaks after pumping the flip-flop (flip-flip) transitions. The Eq.(\ref{eq:fit}) was fitted for the first 11 shells (listed in Table \ref{tab:transition}) around the donor having a significant $\Gamma_{k}$ and $A$, $t_{p}$ and $C$ were used as the fitting parameters. Fitting results are compared with experimental data in Fig. 3 of the main article.